\documentclass[%
 aip,
 amsmath,amssymb,
 reprint%
]{revtex4-1}

\usepackage{graphicx}% Include figure files
\usepackage{dcolumn}% Align table columns on decimal point
\usepackage{color}
\begin{document}

\preprint{AIP/123-QED}

\title{Amplitudes of minima in dynamic conductance spectra of the SNS Andreev contact}
% Force line breaks with \\

\author{Z. Popovi\' c}

 \affiliation{University of Belgrade, Faculty of Physics,
Studentski trg 12, 11001 Belgrade, Serbia}%Lines break automatically or can be forced with \\
\author{S. Kuzmichev}
 \email{kuzmichev@mig.phys.msu.ru}
\affiliation{
Lomonosov Moscow State University, Faculty of Physics, 119991 Moscow, Russia}%
\affiliation{Lebedev Physical Institute, Russian Academy of Sciences, 119991 Moscow, Russia}

\author{T. Kuzmicheva}
\affiliation{%
Lebedev Physical Institute, Russian Academy of Sciences, 119991 Moscow, Russia}%

\date{\today}% It is always \today, today,

\begin{abstract}
Despite several theoretical approaches describing multiple Andreev reflections (MAR) effect in superconductor-normal metal-superconductor (SNS) junction are elaborated, the problem of comprehensive and adequate description of MAR is highly actual. In particular, a broadening parameter $\Gamma$ is still unaccounted at all, whereas a ballistic condition (the mean free path  for inelastic scattering $l$ to the barrier width $d$ ratio) is considered only in the framework of K\"{u}mmel, Gunsenheimer, and Nikolsky (KGN), as well as Gunsenheimer-Zaikin approaches, for an isotropic case and fully-transparent constriction. Nonetheless, an influence of $l/d$ ratio to the dynamic conductance spectrum ($dI/dV$) features remains disregarded, thus being one of the aims of the current work. Our numerical calculations in the framework of an extended KGN approach develop the $l/d$ variation to determine both the number of the Andreev features and their amplitudes in the $dI/dV$ spectrum. We show, in the spectrum of a diffusive SNS junction ($l/d \rightarrow 1$) a suppression of the Andreev excess current, dramatic change in the  current voltage $I(V)$-curve slope at low bias, with only the main harmonic at  $eV=2\Delta$ bias voltage remains well-distinguished in the $dI/dV$-spectrum. Additionally, we attempt to make a first-ever comparison between experimental data for the high-transparency SNS junctions (more than $ 85~\%$)
and theoretical predictions. As a result, we calculate the temperature dependences of amplitudes and areas of Andreev features within the extended KGN approach, which qualitatively agrees with our experimental data obtained using a ``break-junction'' technique.

\end{abstract}

\maketitle
This article may be downloaded for personal use only. Any other use requires prior permission of the author and AIP Publishing. This article appeared in Journal of Applied Physics 128, 013901 (2020) and may be found at https://aip.scitation.org/doi/10.1063/5.0010883
%\begin{quotation}
%The ``lead paragraph'' is encapsulated with the \LaTeX\
%\verb+quotation+ environment and is formatted as a single paragraph before the first section heading.
%(The \verb+quotation+ environment reverts to its usual meaning after the first sectioning command.)
%Note that numbered references are allowed in the lead paragraph.
%
%The lead paragraph will only be found in an article being prepared for the journal \textit{Chaos}.
%\end{quotation}

\section{\label{sec:level1}Introduction}

There is a lack of qualitative comparison between theoretical predictions and experimental practice for the multiple Andreev reflections (MAR) effect manifestation in the dynamic conductance spectrum of superconductor-normal metal-superconductor (SNS) junction. This is especially evident in  ballistic ($l > d$) high-transparent ($85\%< 1-B < 100\%$) type of SNS contact due to several reasons. Here, $l$ is electron mean free path for inelastic scattering, $d$ is the width of thin normal N layer or the constriction, while
$B$ is the probability of the normal reflection from the NS interface and $(1-B)$ is the transmission probability for normal current. The majority of MAR theories \cite{OTBK,Arnold,AverinBardas,Cuevas} do not consider the finite ballistics of the contact along the current direction (hereafter $z$), which could be introduced as $l/d$ ratio, as well as the broadening parameter $\Gamma \equiv \hbar/ \tau$, where $\tau$ is the elastic scattering time of electron. On the other hand, the K\"{u}mmel, Gunsenheimer, and Nicolsky theory (KGN),\cite{KGN} as well as a work by Gunsenheimer  and Zaikin,\cite{zaikin} which nearly reproduces the Octavio, Thinkham, Blonder, and Klapwijk  theory (OTBK) results, \cite{OTBK} consider $l/d$ ratio as an input parameter, instead of taking it infinite.

The simple way to involve a time dependent scattering is an approach, where one can define the probability of the electron to scatter during the MAR process as $1-\exp(-t/\tau)$ that obviously gives zero for the movement time $t\rightarrow 0$ and quickly reaches unity for $t>\tau$. Multiplying both values $t$ and $\tau$ by the Fermi velocity along the current direction ($v_{Fz}$), one can change times to characteristic lengths and roughly estimate the probability to find the ballistic electron in normal metal undergoing $n^{\rm th}$ Andreev reflection as  $P_A(n) =\exp(-n\hspace{1mm} d/l)\hspace{1mm} d/l$. This is because this electron has passed approximately $n \hspace{1mm} d$ distance since the moment of appearance at the NS-interface. At the same time, the probability to find incident carriers along their mean free path is limited by the width of the normal metal, thus $d/l$ (in a simplified case all of them move normally to NS interface).

It is well-known \cite{OTBK,Arnold,AverinBardas,Cuevas,KGN,zaikin} that the $dI/dV$ dynamic conductance spectrum of ballistic high-transparent  SNS contact at any temperatures up to critical $T_c$ exhibits the series of minima at certain bias voltage $eV_n = 2\Delta/n$, where $n$ is the integer number, and $\Delta$ is the superconducting gap. These minima constitute so-called subharmonic gap structure (SGS). In case of absence of the in-gap electron states, a simplified KGN approach to the Andreev current through fully transparent SNS contact could be described as a sum of integrals of electronic density of states (DOS) over $(-\Delta-eV,-\Delta)$ energy range, where $eV$ is a contact bias voltage.\cite{KGN,Gokhfeld} According to the simplified KGN approach, curving of current-voltage characteristics (CVC) at $eV_n$, as well as the shape of the corresponding Andreev minima beneath $dI/dV$ background is defined by the single term of the sum for the Andreev current. This term describes the very last act of the in-gap Andreev reflection, which is very probable in the high-transparent case ($1-B > 85\%$).
This happens for the part of initial incident electron wave packet, and has the running value of the upper limit of integration $(\Delta -(n+1) eV)$. In case of a classical energy dispersion law $E(k)$ and a BCS law of DOS on energy at zero temperatures, infinitesimal probability $B$ and broadening parameter $\Gamma$, this term is proportional to
\begin{eqnarray}\label{eq1}
&&I_A(V)=\\ \nonumber
&&ev_FL_xL_yP_A(n+1)N(0)\int_{-\Delta -eV}^{\Delta-(n+1)eV}\frac{|\varepsilon|}{\sqrt{\varepsilon^2-\Delta^2}}d\varepsilon=\\ \nonumber
&&\frac{V}{R_N}e^{-(n+1)\frac{d}{l}}(\sqrt{1+\frac{2\Delta}{eV}}-\sqrt{(n+1)^2-(n+1)\frac{2\Delta}{eV}}),
\end{eqnarray}
since the normal state resistivity $R_N$ of the SNS contact $R_N = (e^2 v_F (d/l)L_xL_yN(0))^{-1}$, where $L_xL_y$ is the contact area, $d \hspace{1mm} L_xL_y$ is a volume of the normal metal layer, $N(0)$ is normal DOS  and $v_F$  is the Fermi velocity.

For the over-gap bias  $(eV > 2\Delta)$  and direct over-gap current the number $n = 0$, but some outgoing electrons starting their motion through the constriction from the energy range $(-\Delta-eV,\Delta-eV)$, total  $2\Delta$ range, continue to participate in a single Andreev reflection, thus producing Cooper pairs and an additional (excess) current. The electrons starting from $(-\Delta, -\Delta -eV)$ pass toward the conduction band, thus not undergoing Andreev reflection.   In this way  Eq. (\ref{eq1}) transforms to
\begin{equation}\label{eq2}
I_{exc}(eV)=\frac{V}{R_N}(\sqrt{1+\frac{2\Delta}{eV}}-\sqrt{1-\frac{2\Delta}{eV}})e^{-\frac{d}{l}},
\end{equation}
which leading at  high $eV$  to the dependence
\begin{equation}\label{eq3}
I_{exc}(eV >> 2\Delta)=\frac{2\Delta}{eR_N}e^{-\frac{d}{l}}
\end{equation}
of the Andreev excess current $I_{exc}$ on $2\Delta$ (total gap energy range) and the exponent of inverse $l/d$ ratio.
This result for the $I_{exc}$ \cite{KGN,zaikin} deviates from the theories presented in the Refs.  \onlinecite{OTBK} and \onlinecite{AverinBardas} that use full expression for the Andreev reflection probability on energy dependence, by the factor $4/3$ due to the neglecting the over-gap Andreev reflections for $\varepsilon >\Delta$.

Since the part of the Andreev current that gives the main contribution to the curving of CVC at $eV_n$ voltages from Eq. (\ref{eq1}) is proportional to $\exp(-n\hspace{1mm} d/l)$, one should expect a similar dependence of amplitudes of the Andreev minima in $dI/dV$ dynamic conductance spectrum of SNS contact. Whereas it is impossible to get an analytical result for finite temperatures and appearance of the in-gap states in the KGN theory framework, we used the numerical computation to estimate some general tendency of a finite $l/d$ ratio, and the temperature influence on the amplitudes of the Andreev minima and minima area. The latter can be compared with an experimental result of the ``break-junction'' technique,\cite{MorelandEkin} which could produce extremely transparent tunneling constrictions $(85\%<1-B<95\%)$,
while operating with layered superconductors (see for a review Ref. \onlinecite{svetoslav}).
Even then, to make the qualitative comparison between the theory and the experiment, one meets some general difficulties:

1) High but finite transparency of a thin normal metal layer, appearing in break-junction. Even small normal reflection probability presence, let say $B \approx 10\%$,
leading to the decrease of Andreev reflection probability from unity and its dependence on energy, can not be accounted in KGN frameworks. Formally, the effect of $1-B$ decreasing could be roughly taken into account in KGN frameworks  for small B values  ($B < 15\%$) as an additional scattering (thus the decrease of the effective $l/d$ ratio) by modifying exponential multiplier by $\exp(n[\ln(1-B)-d/l])$, which tends to $\exp{(n[-B-d/l])}$ in $B \rightarrow 0$ limit.

2) In the experimental conditions, finite broadening parameter $\Gamma$, which smears CVC features even at $T\rightarrow 0$,
reduces minima amplitude and increases its half-width, depends on the sample homogeneity. At the same time, in accordance with
the KGN theory frameworks, the electron DOS divergence at $|\varepsilon|\rightarrow \Delta$ in the superconducting state leads
to a divergence of the amplitude of $dI/dV$ Andreev minima. Unfortunately, none of MAR effect theories do consider the influence
 of finite $\Gamma$. The theoretical approach to this problem remains an open issue, since one need to consider the loss of carriers
in unoccupated in-gap states (which do exist in this case), in addition to the smearing of DOS distribution on energy. Because the loss
of carriers strongly depends on energy, being maximal just below the gap edges, it makes the analytical description of this process nearly
 impossible for all known models of MAR effect.

3) The ``break-junction'' technique achieves the best results when studying namely layered materials. Note all layered superconductors are non classical ones. They have frequency-dependent superconducting gap $\Delta(\omega)$, the in-gap occupied states at $T\rightarrow 0$ and non-spherical Fermi surface. The latter is not considered in any of the MAR effect theories \cite{OTBK,Arnold,AverinBardas,Cuevas,KGN,zaikin} and needs further theoretical investigation.

\section{Model and methods}

The theoretical model for calculation current voltage characteristic and dynamic conductance, used here is based on formalism previously developed by K\"{u}mmel, Gunsenheimer, and Nicolsky \cite{KGN} and further adopted by Popovi\'c \textit{et al}.\cite{zp1,zp2,zp3,zp4} Details of calculation can be found in these references and we therefore will give  a very brief insight into the formalism used.

\begin{figure}[t]
 \centering
\includegraphics[width=0.5\textwidth]{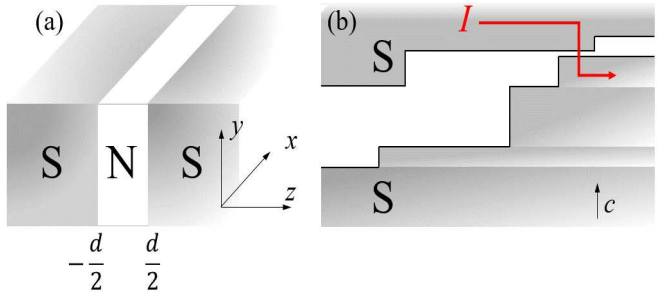}
\caption[]{a) Scheme of the considered SNS junction. b) Scheme of a break-junction formed on steps-and-terraces in a layered sample. Typical terrace width is 20-500\,nm. In such geometry, the current always flows through the constriction along the crystallographic $c$-axis, co-directional with $z$-axis.}\label{sns}
\end{figure}

The theory is applied to the physical situation of two long superconducting electrodes (S) separated by a ballistic normal metal layer (N) of thickness $d$ (Fig. \ref{sns}a). The SNS junction is connected to the voltage source by normal conducting external current leads, so a constant electric field $\mathbf{F}=-\mathbf{e_z}V/d$
($V$ is applied bias voltage) exist only in the N layer.
We use the formalism of the time dependent Bogoliubov-de Gennes equations\cite{KGN,sk} (BdGEs) for electron and hole wave function $u(\mathbf{r},t)$ and $v(\mathbf{r},t)$, respectively, which are
 \begin{eqnarray}
i \hbar \frac{\partial}{\partial
t}u(\mathbf{r},t)&=&\Big[\frac{1}{2m}[\mathbf{p}+\frac{e}{c}\mathbf{A}]^2-\mu\Big]u(\mathbf{r},t)+\nonumber \\
&+&\Delta(z) \Theta(|z|-d/2)v(\mathbf{r},t),\nonumber \\
i \hbar \frac{\partial}{\partial
t}v(\mathbf{r},t)&=&-\Big[\frac{1}{2m}[\mathbf{p}-\frac{e}{c}\mathbf{A}]^2-\mu\Big]v(\mathbf{r},t)+\nonumber \\
&+&\Delta(z) \Theta(|z|-d/2)u(\mathbf{r},t), \label{bdge}
\end{eqnarray}
where $\Theta (z)$ stands for the Heaviside step function, $\mu$ is the chemical potential, while the temperature dependent vector potential is $\mathbf{A}=\mathbf{e}_zcVt/d\Theta(d/2-|z|)$. This equations are combined with the relaxation time model for charge transport. Through this model $l$  enters into calculation. This implies $l>d$ for the N layer.

The KGN theory frameworks  were extended by us earlier to introduce the anisotropy of the superconducting gap, as well as some effect of ferromagnet barrier on Andreev transport.\cite{zp1,zp2,zp3,zp4} We approximate the spatial variation of the pair potential by step function  $\Delta(\theta)\Theta(|z|-d/2)$, where $\Delta(\theta)=\Delta_{max}(1+0.5A(\cos(4\pi\theta)-1))$ reflects anisotropy of the order parameter $\Delta$ in the $k_{xy}$ momentum space (corresponds to the $ab$-plane of the real space), such that $\tan(\theta)=k_{y}/k_{x}$. \cite{svetoslav}
The coefficient $A$ reflects the gap anisotropy in percentages, while $\Delta_{max}$ is the maximum amplitude. Here we use some minor amount of the anisotropy ($A=2 \%$)
to get rid of the $dI/dV$ divergence at the Andreev minima positions $eV_n=2\Delta/n$. For a pure s-wave case of superconducting electrodes the coefficient $A$ is equal zero, and  $\Delta_{max}$ is the bulk superconducting gap $\Delta$. The temperature dependence of $\Delta$ is given  approximately by $\Delta(T)=\Delta(0)\tanh(1.74\sqrt{T_c/T-1})$.

The solutions of time dependent BdGEs in the normal metal barrier are given in the form of the quasiparticle wave packets  $u_k^{\pm}$ and $v_k^{\pm}$  which move in the electric field due to the applied voltage $V$. This solutions are used to calculate average current density which in the relaxation-time model is given by\cite{KGN,zp1}
\begin{eqnarray}
\langle \mathbf{j}\rangle =
-\frac{e}{2m}\hat{\sum_k}\Big\{f_0(E_k)\Big[\langle
u_{k}^{+*}\mathbf{P}u_{k}^{+}\rangle + \langle
u_{k}^{-*}\mathbf{P}u_{k}^{-}\rangle\Big] +\\ \nonumber
+(1-f_0(E_k))\Big[\langle
v_{k}^{+}\mathbf{P}v_{k}^{+*}\rangle + \langle
v_{k}^{-}\mathbf{P}v_{k}^{-*}\rangle\Big] \Big\}.\label{struja}
\end{eqnarray}
Here $f_0$ is the Fermi distribution function and
$\mathbf{P}=[-i\hbar \mathbf{\nabla}+e\mathbf{A}/c]$ is the
gauge-invariant momentum operator. The averaged momentum densities $\langle u_{k}^{\pm *}\mathbf{P}u_{k}^{\pm}\rangle$ and $\langle v_{k}^{\pm}\mathbf{P}v_{k}^{\pm *}\rangle$ are calculated in the same way as in Refs. \onlinecite{KGN} and \onlinecite{zp1,zp2,zp3,zp4} and they are proportional to corresponding multiple Andreev reflection probability amplitudes
\begin{eqnarray}
 \langle u_{k}^{\pm *}\mathbf{P}u_{k}^{\pm}\rangle \approx
\sum_{n=0}^{\infty}\Big|A_{2n}^{\pm}(E\pm\frac{eV}{2})\Big|^2,\label{upu1}\\
 \langle v_{k}^{\pm}\mathbf{P}v_{k}^{\pm *}\rangle \approx
\sum_{n=0}^{\infty}\Big|A_{2n+1}^{\pm}(E\pm\frac{eV}{2})\Big|^2,\label{upu2}
\end{eqnarray}
where $A_{2n}^{\pm}(E)=\prod_{\nu=1}^{2n}\gamma (E\pm\nu eV\mp\frac{eV}{2})$, $A_{2n+1}^{\pm}(E)=\prod_{\nu=1}^{2n+1}\gamma (E\pm\nu eV\mp\frac{eV}{2})$. So,  $A_{2n}^{\pm}$   $A_{2n+1}^{\pm}$ are the probability amplitude that a quasiparticle at energy $E$ starts to move as an electron against (+) or opposite to (-) the field will reappear in the normal metal barrier as an electron after $2n$ Andreev reflections and as a hole after $2n+1$ Andreev reflections, respectively. Therefore, $|\gamma(E)|^2$ is the probability that an AR occurs at energy $E$ in the phase boundary of a semi-infinite superconducting electrode and $\gamma=(E-i(\Delta^2-E^2)^{1/2})/\Delta$ for $E<\Delta$ (whereas for $E>\Delta$ it may be approximate by zero as in Ref. \onlinecite{KGN}).

Note that, the solution of time dependent BdGEs  in the limit of vanishing voltage must turn into the stationary quasiparticle wave function of an superconductor/normal metal junction. In the theory used here \cite{KGN,zp1,zp2,zp3,zp4} voltage dependent solutions evolve from the bound states (while in Ref. \onlinecite{zaikin} the solutions evolving from the scattering states play the dominant role) and quasiparticle start their motion in the electric field from energy $|E|<\Delta$ after each Andreev reflection.\cite{kumel} Since, the SNS junction is connected by normal conducting external current leads to the voltage source almost all  quasiparticle excitations have energies $|E|<\Delta$ and they are completely Andreev reflected at the external interfaces between the leads and the junction.

After very extensive calculations presented in detail in Refs. \onlinecite{KGN} and \onlinecite{zp1} it can be obtained that the total current density $\langle \mathbf{j}\rangle$ (which has only $z$ component) is the sum of Ohmic current density $\langle\mathbf{j_N}\rangle$ and current density due to Andreev reflection $\langle \mathbf{j_{AR}}\rangle$. In the following we calculate the total current $I$ (and corresponding conductance $dI/dV$) which is connected to the total current density $\langle j\rangle_z$ via cross section area $L_xL_y$
\begin{equation}
I=I_N+I_{AR}=L_xL_y(\langle j_N\rangle_z +\langle j_{AR}\rangle_z).
\end{equation}
The total current is normalized by the temperature dependent current $I_0=2\Delta(T)/(eR_N)$.

Experimentally, in order to make SNS junctions for Andreev spectroscopy studies, we used a break-junction technique.\cite{MorelandEkin} Its specialities, some details and discussions could be found elsewhere.\cite{svetoslav} The sample prepared as a thin rectangular plate with dimensions about $3 \times 1.5 \times 0.2$\,mm$^3$ was attached to a springy sample holder by four-contact pads made of In-Ga paste at room temperature. After cooling down to $T = 4.2$\,K, the holder was gently curved, thus cracking the bulk sample, with a formation of two clean cryogenic surfaces separated with a weak link (see Fig.~\ref{sns}(b)), a kind of ScS contact where $c$ is a constriction. The resulting constriction turns far from potential and current leads, which prevents junction overheating and provides true four-point probe.
In magnesium diborides,\cite{MgB2 1,MgB2 2,svetoslav,MB2004JetpL,MB2004SSC} the used technique provides constrictions with various transparency.

Under fine tuning the curvature of the sample holder, the two cryogenic clefts slide apart touching onto various terraces. Such tuning enables to adjust the constriction area in order to realize a desired tunneling SNS regime. During the experiment, the cryogenic surfaces remain tightly connected when sliding that prevents impurity penetration into the crack and protects the purity of cryogenic clefts.

In a layered sample the crack naturally splits the $ab$-planes, with a formation of steps and terraces. The height of the step is usually about $(10-100)c$ unit cell parameters, whereas the typical terrace size appears about $20-500$\,nm. These features, as well as the geometry of the break-junction setup are presented in Fig.~\ref{sns}(b). While the constriction area ``c'' of planar ScS-contact in $ab$ crystallographic plane can be of any shape, its area variation under the fine tuning defines just normal resistance $R_N$. The current passes along $c$-direction, and namely the junction width $d$ defines ballistic $l/d$ ratio (note, $l$ is inelastic scattering length along $c$-direction, or $z$-axis in Fig.~\ref{sns}). Typically this is the case for polycrystalline sample of layered compound as well\cite{EPL,SmPRB,svetoslav,Nd}.

Since the constriction appears as a part of a terrace (see Fig.~1(b)), the typical in-plane dimensions of the break-junction $L_x,L_y$ supposed to be much more than the Cooper pair size $\xi_0$. Since the macroscopic scale (which could be estimated on accounting relatively low resistance for some contacts obtained), there is no need in accounting of the dimensional quantization effect. Despite the impossibility of direct lengths measuring or visualization of the break-junction obtained, it can be assumed that along the $c$-direction, a couple of distorted (``broken'') layers of crystal structure act as the constriction separating the intact superconducting banks. Then, the effective constriction width $d$ is compared with $(2-3)c$ unit cell parameters.

Here we present the experimental data for the ballistic high-transparency barriers ($95\% \textendash 98 \%$),
which are electrically equivalent to a thin layer of normal metal (SNS) with the thickness $d$ about the superconducting coherence length  $\xi_0$. In the majority of the break junctions in Fe-based superconductors we studied, the resulting constriction formally act as normal metal, with the $I(V)$ and $dI(V)/dV$ typical for the clean classical SNS junction.\cite{OTBK,Arnold,AverinBardas,Cuevas,KGN,zaikin} We consider the presence of the excess current and clearly visible SGS, as well as their disappearing over $T_c$, as the benchmark of the MAR regime developing and the ballistic character of our break-junctions. As mentioned above, using mechanical readjustment, it is possible to produce contacts of various geometry and area. As a common practice, considered are only the break-junctions which SGS position do not depend on the contact in-plane geometry (as an example, see Figs.~9,~12,~16 in review \cite{svetoslav}), and which CVC is symmetric above and below $T_c$.

In our studies, the dynamic conductance spectra were measured directly by a standard modulation technique.\cite{LOFA,svetoslav} We used a current source with an admixture of $ac$ frequency about 1\,kHz from the external oscillator. The results obtained with this kind of setup are insensitive to the presence of parallel ohmic conduction paths; if any path is present, the dynamic conductance curve shifts along the vertical axis only, while the bias stay unchanged. As a result, the ``break-junction'' technique is a precise and high-resolution local probe of the superconducting order parameter, its temperature dependence and a fine structure.\cite{svetoslav}

\section{Results and Discussion}

\begin{figure}[b]
 \centering
\includegraphics[width=7.6cm]{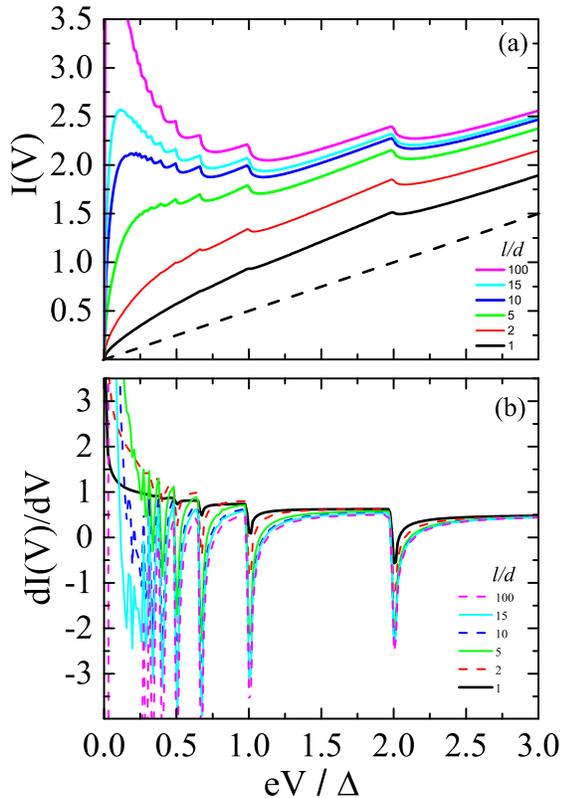}
\caption[]{The numerical result for  (a) CVCs $I(V)$  and (b) corresponding dynamic conductance spectra $dI(V)/dV$, for the \textit{fully transparent} SNS contact ($T\rightarrow 0$) with the $l/d$ ratio variation, calculated in the framework of KGN theory.\cite{KGN} The subharmonic gap structure (SGS) is well visible in all CVCs and conductances. The location of the SGS corresponds to  $eV_n = 2\Delta/n$, where $n$ is integer number, $\Delta$ is the superconducting gap amplitude. Normal state ohmic dependence ($T > T_c$) is shown by dashed line in panel (a).}\label{cvc}
\end{figure}

The numerical results of CVC calculation with the $l/d$ ratio variation ($T\rightarrow 0$), according to the KGN theory \cite{KGN} which accounts for a realistic angle dependent three-dimensional geometry of the contact, are presented in Fig. \ref{cvc}(a).  Note, the bias voltage is normalized to superconducting gap value $\Delta$. One can say, this $l/d$ ratio demonstrates the ballistic ``quality'' of a microcontact. Since KGN approach is valid only for fully transparent constrictions ($B = 0$), contacts with the largest $l/d$ ratio are in deep ballistic limit. In the latter case, KGN theory predicts the extreme rise of the Andreev current at low biases (so-called ``foot'' region), as it could be seen for three upper CVCs (violet, cyan and blue) in Fig.~\ref{cvc}(a). The range of $l/d$ values, which produces the situation, when the current at the edge of the foot structure (``the edge current'') overcomes a total current of the  SNS-contact at $\Delta/2 < eV < 2\Delta$, is limited by 7-9. According to the $\exp[n\ln{(1-B)}]=(1-B)^n$ correction to be made for the absolute ballistics $l/d\rightarrow \infty$  case (see the $1^{st}$ point of the list of difficulties presented above), this limit roughly corresponds to the normal reflection probability  $B\approx 12\%$
in case of finite contact transparency. In other words, the CVC of the SNS-contact having  $1-B\approx 88\%$
transparency, even absolutely ballistic, will not demonstrate any drastic rise of current in foot region, which, however, agrees with Averin-Bardas\cite{AverinBardas} theory predictions for the case of high transparency (more than $80\%$)
constriction. This fact is in good agreement with our experimental findings.  In general, Fig.~\ref{cvc}(a) shows a tendency of dramatic flattering of the CVC slope in the foot region with $l/d$ or $(1-B)$ decrease. As for high bias voltages $eV \gg 2\Delta$, there observed Andreev excess current is proportional to $\exp[-d/l]$, thus corresponding to Eq.~(\ref{eq3}).

All $I(V)$ characteristics from Fig. \ref{cvc}(a), as well as their derivatives  $dI/dV$ from Fig. \ref{cvc}(b), show well visible subharmonic gap structure  at $eV_n = 2\Delta/n$, where $n$ is the integer number. It was shown,\cite{OTBK,KGN,zaikin} the SGS position is determined by this formula at any temperatures up to $T_c$. As one can see from the numerical calculations of dynamic conductance spectra presented in Fig. \ref{cvc}(b), it is possible to make the same conclusion for the $l/d$ ratio variation. This variation changes a dynamic conductance background (especially for biases $eV < \Delta$), as well as the amplitudes and areas of the Andreev minima, but does not change their locations.

The divergence of BCS density of states near the gap edge leads to the divergence of minima amplitudes, resulting in KGN frameworks. To overcome such the theoretical divergence, we used as small as 2\%
anisotropy for the superconducting $s$-wave order parameter distortion. The visible amplitude of the $n = 1$ minima (``$2\Delta$'' feature) in dynamic conductance spectra (Fig.~\ref{cvc}(b)) dominates over other minima only for black and red curves with $l/d \leq 2$, and the number of a detectable minima is limited there, even we consider the case of $T\rightarrow 0$ and $\Gamma \rightarrow 0$. The flattering of the foot region is well visible in this case. The usage of finite bias step in Figs.~2 (a) and (b) leads to an averaging of beating behaviour of the dynamic conductance at small bias voltage $eV \ll \Delta$ (see Appendix~A for details and some technical issues).

\begin{figure}[b]
 \centering
\includegraphics[width=7.6cm]{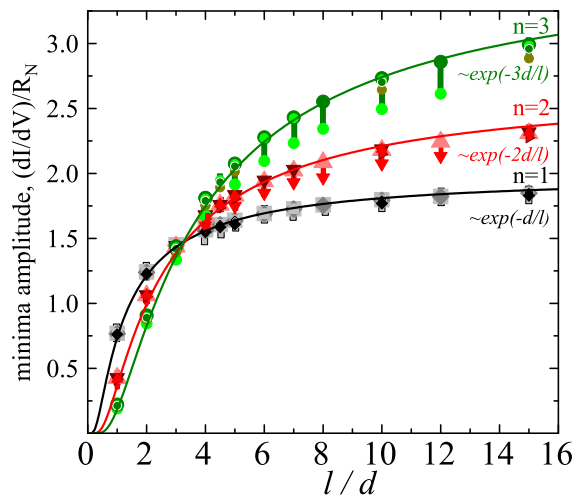}
\caption{Visible amplitudes of  SNS Andreev dynamic conductance minima for $n = 1,2,3$  as a function of $l/d$ ratio (black squares, red triangles, green circles, correspondingly) at $T\rightarrow 0$  in KGN framework (numeric estimation). The contact diameter variation range is $d =(0.2-3)\xi_0$. Points of the same hue, but different shade demonstrate the range of reproducibility for the finite $dV$-step of numerical calculations. Lines demonstrate $\exp(-n ~d/l)$ dependencies (is shown with the corresponding color).}
\label{Ald}
\end{figure}

Figure \ref{Ald} presents the results for the numerical estimation of amplitude of the $dI/dV$ Andreev minima of fully transparent SNS contact (having contact diameter range $d = (0.2 - 3)\xi_0$, which ensure that no one Andreev in-gap level is visible) at  $T\rightarrow 0$ with the $l/d$ ratio variation. Note that, $l$ and $d$ are normalized by coherence length $\xi_0$, and $n$ are natural numbers, which define a location of the SGS minima $eV_n = 2\Delta/n$. Vertical bars depict the numerical uncertainty of the result. Formally, the KGN theory does not consider any inelastic processes for the $l/d > 1$, presented here, and all electrons and holes pass normal metal ballistic. It is clear from Fig. \ref{Ald}, the considered $l/d$ range is separated by two parts at $l/d\approx 3$, where all Andreev minima comprising the subharmonic gap structure have nearly the same visible amplitude. For $l/d < 3$  one approaches the regime of the mixed (elastic and thermal) transport through the SNS-contact. In this case an amplitude of the Andreev minima corresponding to $n = 1$ dominates over all other minima of SGS. The opposite situation could be seen for the high-quality ballistic junction with $l/d > 3$. In real constriction having finite transparency, this limiting value could be higher and approaches, for example, $(l/d)_{\rm lim} = 4-5$ for $1-B\approx  90\%$ normal transmission (according to the $\exp(n[\ln(1-B)-d/l])$ estimation from the 1$^{\rm st}$ point of the list of difficulties presented above).

It is possible to make an analytical approximation for the numerical result presented in Fig. \ref{Ald} for $T \rightarrow 0$. If we denote an amplitude of the $n^{\rm th}$ Andreev minima for infinite $l/d$ ratio (absolute ballistic constriction) as $A_n(\infty)$, then we can expect the $A_n(l/d) = A_n(\infty)\exp(-n\hspace{1mm} d/l)$ dependence (plotted as lines in Fig. \ref{Ald}) according to the expression for the current   Eq. (\ref{eq1}). Thus, at $B\rightarrow 0$ and $T\rightarrow 0$ the amplitude ratio of any adjacent Andreev minima could be derived as $r_A(l/d)\equiv A_{n+1} / A_n = r_A(\infty)\exp(-d/l)$.

\begin{figure}
 \centering
\includegraphics[width=7.6cm]{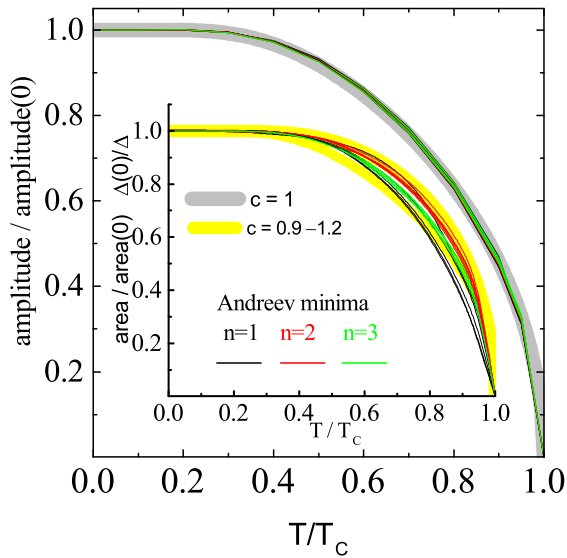}
\caption{Temperature dependence of Andreev minima amplitude of $n = 1, 2, 3$ Andreev dips numeric estimation (thin black, red, green lines, correspondingly)  normalized to their magnitudes at $T\rightarrow 0$. Comparison with the $\tanh[\Delta/(2k_BT)]$ dependence is shown by bold gray arc curve. \textbf{The inset} shows temperature dependence of the area of $n = 1,~2,~3$ Andreev dips numerical estimation (thin lines of the corresponding colour), normalized to $\Delta(T)$, as well as to their $area(0)/ \Delta(0)$ value. Comparison with $\tanh[c \Delta/(2k_BT)]$ function with the free parameter $c = 0.9-1.2$ is shown by bold yellow lines. Both data sets are numerically estimated in KGN frameworks for the contact with $d = (0.2 - 3) \xi_0$ and $l/d = 3/2 - 100$ variation.}
\label{At}
\end{figure}

The temperature dependence of Andreev minima amplitude $A_n(T)$ in KGN theory frameworks was checked for the variety of normal metal thicknesses (the same $d = (0.2 - 3)\xi_0)$ and ballistic quality of the constriction $(3/2 < l/d < 100)$,  see Fig.~\ref{At}. In all cases the $A_n(T) / A_n(0)$ results for the first ($n = 1$), second ($n=2$) and third ($n=3$) minima are qualitatively the same (see thin curves in Fig.~\ref{At}, where their colors represent the SGS number ``$n$''). The scattering range of our numerical estimation is presented in Fig.~\ref{At} by bold gray arc tending to $\tanh[\Delta/(2k_BT)]$, which temperature dependence is close to the BCS superconducting $\Delta(T)$. Note that in this work all the amplitudes are initially normalized to $2\Delta(T)$ value, since the current is normalized to $I_0 = 2\Delta(T) / (eR_N)$. The amplitude residual is shown in Fig. \ref{Atrez} of Appendix B.

The inset of Fig. \ref{At} demonstrates the temperature trends of Andreev minima areas normalized to their $T=0$ value for the same numerical estimations in the range of normal metal width $d$ up to $3 \xi_0$, limiting the visual appearance of 1$^{\rm st}$ Andreev bound state, as separate CVC feature. As an envelope curve for $dI(V)/dV$ spectra, we used a piecewise function constructed as a series of straight line segments, which connect all points with maximal dynamic conductivity at $eV \rightarrow 2\Delta/n$ from the left. Being the easiest solution, such the piecewise function produces a background uncertainty. The area is estimated between the $dI(V)/dV$ spectra and its envelope curve. For details, see the second paragraph of Appendix B.

All the numerical results lay within bold yellow region (see the inset of Fig. \ref{At}), which is constructed by the $\tanh[c \Delta/(2k_BT)]$ function with the variation of a free parameter $c = 0.9-1.2$. The results for $n=1$ minima (black thin lines) slightly differ to those for $n=2,~3$, but can be roughly approximated by the same $\tanh[c \Delta/(2k_BT)]$ function.

\begin{figure}[t]
 \centering
\includegraphics[width=7.6cm]{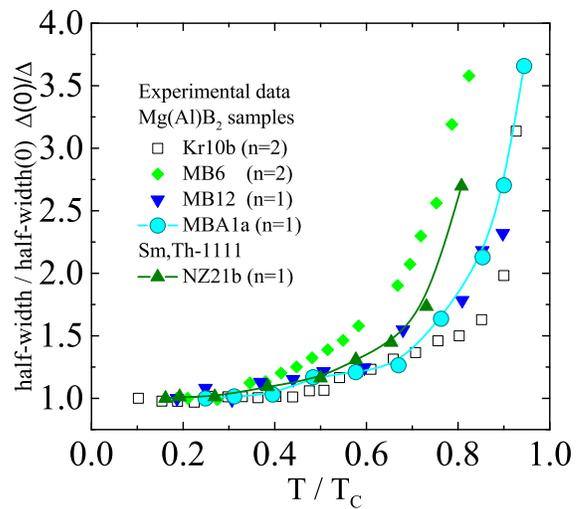}
\caption{Temperature dependence of relative half-width of Andreev minima normalized to their magnitudes at $T\rightarrow 0$, as well as $\Delta(T)$. Experimental results for Mg(Al)B$_2$ and Sm-1111 superconductors.}
\label{HW}
\end{figure}

\begin{figure}[t]
 \centering
\includegraphics[width=7.6cm]{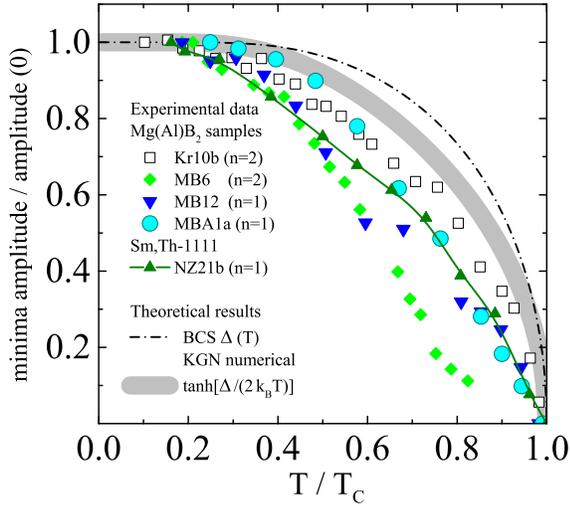}
\caption{ Temperature dependence of Andreev minima amplitudes normalized to those at $T\rightarrow 0$. Bold gray curve demonstrates the range of reproducibility for $d = (0.2-3)\xi_0$ according the KGN theory (data is taken from Fig.~\ref{At}; no broadening parameter $\Gamma$ is accounted). Experimental results for Mg(Al)B$_2$ and Sm-1111 superconductors are shown by points. Weak-coupling BCS theory result for $\Delta(T)$ is presented by dash-dotted curve for comparison.}
\label{Aexp}
\end{figure}

\begin{figure}[t]
 \centering
\includegraphics[width=7.6cm]{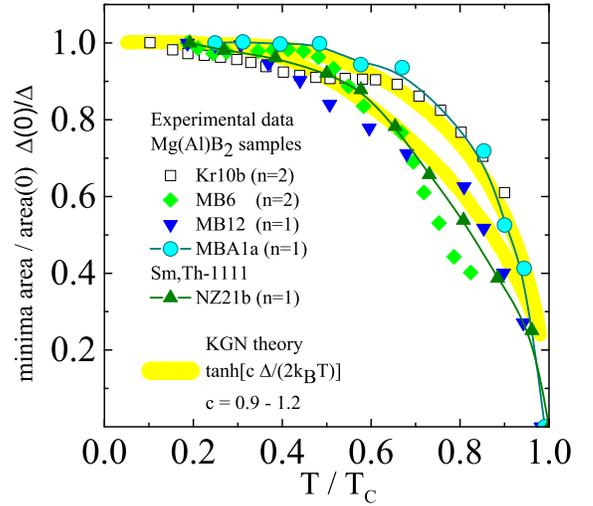}
\caption{ Temperature dependence of Andreev minima areas normalized to their magnitudes at $T\rightarrow 0$, as well as $\Delta(T)$. The range of the numerical results in the framework of KGN theory are presented by bold yellow lines. Experimental results for $Mg(Al)B_2$ and Sm-1111 superconductors are shown by points.}
\label{Areaexp}
\end{figure}

Here we use the experimental data obtained for break-junctions formed in MgB$_2$, MgB$_2$ + MgO, Mg$_{1-x}$Al$_x$B$_2$ superconducting samples, as well as in Sm$_{1-x}$Th$_x$OFeAs iron-based superconductor (so called 1111). For these materials, the in-plane $\xi_0$ is as low as several unit cell parameters, therefore one could roughly estimate the constriction width $d \approx \xi_0$. Being two-gap superconductors, magnesium diborides as well as the 1111 system are expected to have isotopic s-wave order parameters, which is important to get rid of the uncertainty that produces any gap anisotropy. SNS Andreev spectra showing two distinct SGS (caused by the $\Delta_L$ and $\Delta_S$ gaps) for several break-junctions were recorded by us earlier with a variation of $T$.\cite{MgB2 1,MgB2 2,SmPRB} These spectra were selected to plot temperature dependence of half-widths, amplitudes, and areas for $n = 1,2$ Andreev minima (presented by symbols in Figs.~\ref{HW},~\ref{Aexp},~\ref{Areaexp}, respectively). Since these superconductors have large $\Delta_L/\Delta_S$ ratio (about 3-6), the two SGS are non-overlapped, thus facilitating the observation of undistorted Andreev minima. With it, SGS for the small $\Delta_S$ gap is usually located in the area of the drastic rise of the dynamic conductance at low biases (``foot''), which makes the amplitude and half-width estimation ambiguous. In order to exclude that factor, we use the large gap SGS to made such estimate.

The temperature dependence of the half-width of the Andreev minima (various $n$ numbers) normalized to both, its zero value and $\Delta_L(T)/\Delta_L(0)$ are shown on Fig. \ref{HW}. Results  were  obtained in several superconducting samples. All the data plotted here was estimated directly from the dips in our experimental spectra and was not averaged. Noteworthily, the dependence showed by olive up triangles and blue down triangles similarly evolve with temperature, up to $\sim T_c/2$, as one can see in Fig.~\ref{HW} as well as in Fig.~\ref{Aexp} showing temperature dependence of Andreev minima amplitude. This demonstrate the temperature trend of their relative  minima amplitudes, despite these data correspond to the behavior of Andreev minima for the superconductors of different families (MgB$_2$ and Fe-based 1111). The slowest growing dependence (open black and cyan symbols) in Fig. \ref{HW} seem corresponding to the most homogeneous contact points, with, therefore, the minimum broadening $\Gamma$. The latter obviously makes such data an ideal candidate to compare with MAR theories, those considering zero $\Gamma$.

The data range shown in Figs.~\ref{Aexp}  and \ref{Areaexp} by solid points represents the experimental uncertainty. For the most qualitative SNS-contacts on magnesium diborides (black and cyan points), the data keep close to the variety of numerically obtained data using the extended KGN theory (bold arches). The slightly averaged experimental dependence of amplitudes $A_n(T)/A_n(0)$  are strain after the theoretical predictions (bold gray line in Fig.~\ref{Aexp}). The deviation from this trend is apparently caused more by the changes in the scattering rate than the subharmonic order $n$.  This means the larger $\Gamma$, the lower the temperature dependence of the relative minima amplitude. It seems, the experimental results tends to $\tanh[\Delta/(2k_BT)]$ function, rather than standard BCS theory result $\Delta(T)$ (dash-dot curve). All experimental data plotted in Fig. \ref{Aexp} demonstrate the decrease of the Andreev minima amplitudes starting from the lowest temperatures, in accordance with OTBK and KGN theories.\cite{OTBK,KGN}

The results for the relative area of the Andreev minima estimation from our experimental data for the dynamic conductance of SNS contacts in Mg(Al)B$_2$ superconductor show some dispersion due to the distinct level of scattering (see points in Fig.~\ref{Areaexp}). The resulting dependence of relative area on $T$ for green rhombs ($n=2$ minima) shows the signs of some kind of a ``tail'' that tends to $T_c$ starting from $T/T_c = 2/3$. Taking into account that the corresponding dependences in Figs.~\ref{HW}  and \ref{Aexp} for this SNS contact (green rhombs) deviates significantly from the set of all others, one can conclude, it has the shortest smearing time, thus the strongest $\Gamma$, which can result in the loosing of states and the suppression of this Andreev dip well before the $T_c$. In case when $\Gamma$ value is comparable to $\Delta$, the number of empty states appears inside the superconducting gap region, and electrons involved in MAR process may be lost in the in-gap range of energies, thus decreasing the Andreev part of the current, as well as the amplitude and area of the corresponding Andreev minima. However the variety of these experimental data points correspond to the predictions of the KGN theory well (the range limited by bold yellow curves) and can be roughly approximated by the $\tanh[c \Delta/(2k_BT)]$ function having the only free parameter $c = 0.7 - 1.4$.

\section{Conclusions}

In conclusion, we made qualitative comparison between theoretical predictions of the extended KGN theory \cite{KGN,zp1,zp2,zp3}  in isotropic case and experimental data by the ``break-junction'' technique \cite{svetoslav} for areas and amplitudes $A_n$ of the Andreev minima in the dynamic conductance spectra of ballistic high-transparent $(B < 15\%)$
superconductor - thin normal metal - superconductor (SNS) Andreev contacts on temperature. Since the KGN theory do not consider finite broadening parameter $\Gamma$, we get the best accordance for those break-junctions supposed to have the highest level of homogeneity (due to the slow increase of the half-width of Andreev minima with $T$) and, thus, the pretty small $\Gamma$.

Our experimental $A_n(T)$ dependences do not show any region of the minima amplitude increase with temperature, and tend to the theoretical one $A(T) \sim \Delta(T) \tanh[\Delta/(2k_BT)]$. The temperature dependences of the minima area correspond to the KGN theoretical predictions well, and can be qualitatively described with tanh-like function with the single free parameter.

We estimated, how the minima amplitudes in the dynamic conductance of SNS contact depend on the mean free path to the constriction width ratio $A_n(l/d)$ at $T\rightarrow 0$ and checked the $\exp(-n\hspace{1mm} d/l)$ analytical approximation fits the numerical result. We have shown that the amplitude ratio of any adjacent Andreev minima  is $r_A(l/d)\equiv A_{n+1} / A_n = r_A(\infty) exp(-d/l)$ at $T \rightarrow 0$.

\begin{acknowledgments}
We are grateful to S. I. Krasnosvobodtsev, as well as L. G. Sevastyanova, K. P. Burdina, V. K. Gentchel, B. M. Bulychev, and N. D. Zhigadlo for the samples provided at our disposal. The work of S.A.K. ware supported by RFBR project 18-02-01075a. T.E.K. acknowledges the state assignment of the Ministry of Science and Higher Education of the Russian Federation (topic ``Physics of high-temperature superconductors and novel quantum materials'', No.  0023-2019-0005). The work of Z. P. was supported by Serbian Ministry of Education, Science and Technological Development, Project No. 171033.
\end{acknowledgments}

\section*{DATA AVAILABILITY}
The data that support the findings of this study are available from the corresponding author upon reasonable request.

\section{Appendix A}
To overcome the issue with vanishing broadening parameter $\Gamma$ (the $2^{nd}$ point of the list of difficulties presented above) leading to some divergences in the theory from one hand, and finite $\Gamma$ that smears features of experimental dynamic conductance spectra, from the other hand, we use the finite $dV$-step for the partial DOS on energy, as well as for CVC calculations. The results of calculation obtained for various normal metal width $d$, distinct $l/d$ ratio, have shown the counterintuitive enhancement of the numerical reproducibility with an increase in the $dV$-step of calculation up to $0.01\Delta$, due to the limiting of $dI/dV$ minima amplitudes and the reproducibility of its shapes. The other numerical way to overcome the minima divergences is to extend the KGN model and involve some superconducting gap anisotropy in the $k$-space.\cite{zp1,zp2,zp3,zp4} We used $2 \%$
anisotropy and standard four-fold superconducting s-wave order parameter distribution.

The absence of well-defined minima series in a foot region of dynamic conductance spectra for $l/d = 10$ to $15$ (blue and cyan curves in Fig~2(b), respectively) comes from the numerical limiting of the bias step  ($dV = 0.01 \Delta$ is constant), while the minima width decrease as $2/n - 2/(n+1) = 2 / n(n+1)  \approx (eV)^2$, since $n$ is an integer part of $2\Delta/eV$. To overcome this technical issue one need to increase the density of calculated points as $1/(eV)^2$.

\section{Appendix B}

\begin{figure}[t]
 \centering
\includegraphics[width=7.6cm]{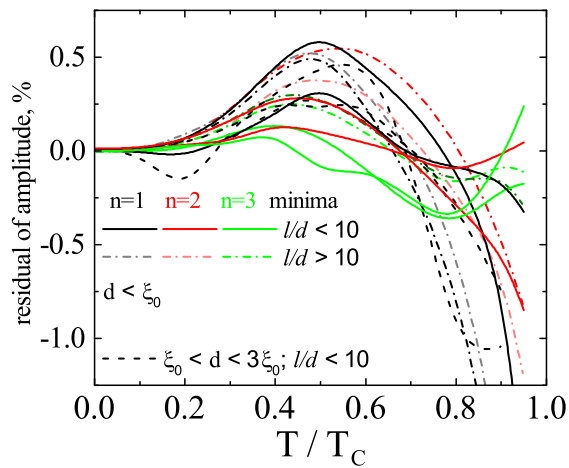}
\caption{ Temperature dependence of amplitude residual in percents for all curves from Fig.~\ref{At} (shows better then 1\% correspondence).}
\label{Atrez}
\end{figure}

Theoretical dependence of Andreev minima amplitude versus temperature normalized to their values at $T\rightarrow 0$, $A_n(T)/A_n(0)$,  in Fig.~4 find quantitative coincidence. This is presented in Fig.~\ref{Atrez} where their amplitude residual in comparison with $\tanh[\Delta / (2k_BT)]$ function is shown. As an example see solid and dashed lines that represent residual to one and the same function in this figure. As one can see at Fig.~\ref{Atrez} the residual between them for all curves from Fig.~\ref{At} does not exceed $\approx 1\%$,
thus meeting good quantitative correspondence. The worst coincidence with analytical approximation we get for $n=1$ Andreev minima, similarly for relative areas of minima on temperature (from the inset of Fig.~\ref{At}). The best one is obtained for $n=3$ (error $< 0.5$\%).
In any case, the region near $T_c$ produces the largest errors, since we need to compare the values tending to zero. One can conclude from Fig.~\ref{Atrez} that both $l/d$ and $d/\xi_0$ variation has no influence on the correspondence quality between amplitude $A_n(T)$ dependence and $\tanh[\Delta / (2k_BT)]$ from Fig.~\ref{At}.

For the estimation of an area of the first $n=1$ minima (presented by thin black lines in the inset of Fig.~4) we used the upper limit $3\Delta$, in order to make its integration range the same as for $n=2$ minima, instead of infinite. An accurate estimation of this area is a difficult task both for theory and for experiment, since the $dI/dV$ background uncertainty.

\nocite{*}
\bibliography{aipsamp}% Produces the bibliography via BibTeX.

\begin{thebibliography}{999}

\bibitem{OTBK} M. Octavio, M. Tinkham, G. E. Blonder, and T. M. Klapwijk, Phys. Rev. B {\bf 27}, 6739 (1983).
\bibitem{Arnold} G. B. Arnold, J. Low Temp. Phys. {\bf 68}, 1 (1987).
\bibitem{AverinBardas} D. Averin and A. Bardas, Phys. Rev. Lett. {\bf 75}, 1831 (1995).
\bibitem{Cuevas} J. C. Cuevas, A. Mart\'{\i}n-Rodero, and A. Levy Yeyaty, Phys. Rev. B {\bf 54}, 7366 (1996); A. Poenicke, J. C. Cuevas, and M. Fogelstr{\o}m, ibid. {\bf 65}, 220510(R) (2002).
\bibitem {KGN} R. K$\ddot{u}$mmel, U. Gunsenheimer, and R. Nicolsky, Phys. Rev. B {\bf 42} 3992  (1990).
\bibitem{zaikin}
U. Gunsenheimer and A.D. Zaikin, Phys. Rev. B {\bf 50}, 6317  (1994).
\bibitem{Gokhfeld} D. M. Gokhfeld, Supercond. Sci. Technol. \textbf{20}, 62-66 (2007).
\bibitem{MorelandEkin} J. Moreland and J. W. Ekin, Appl. Phys. Lett. \textbf{47}, 175 (1985); J. Moreland and J. W. Ekin, J. Appl. Phys. \textbf{58}, 3888 (1985).
\bibitem{svetoslav}
S. A. Kuzmichev and T.E. Kuzmicheva, Low Temp. Phys. {\bf 42} 1008 (2016) [Fiz. Nizk. Temp. {\bf 42}, 1284 (2016)].

\bibitem {zp1} Z. Popovi\' {c},   L. Dobrosavljevi\' {c} - Gruji\' {c}, and R. Zikic,   Phys. Rev. B {\bf 85}  174510  (2012).

\bibitem {zp2} Z. Popovi\'{c}, L. Dobrosavljevi\'{c}-Gruji\'{c}, and R. Zikic,
  J.  Phys. Soc. Jpn.  {\bf 82}, 114714 (2013).

\bibitem{zp3}	
Z. Popovi\' {c}, R. Zikic, and L. Dobrosavljevi\' {c}-Gruji\' {c},
Prog. Theor.  Exp. Phys. {\bf 2015}, 103I01  (2015).

\bibitem{zp4}
Z. Popovi\' {c}, P. Miranovi\' {c}, and R. Zikic,
 Phys. Status Solidi b {\bf 255}, 1700554  (2018).

\bibitem{sk}
 R. K$\ddot{u}$mmel and W. Senftinger, Z. Phys. B {\bf 59}, 275 (1985).

\bibitem{kumel}
A. Jacobs, R. K$\ddot{u}$mmel, and H. Plehn, Superlattices and Microstructures {\bf 25}, 669  (1999).



\bibitem{MgB2 1} S. A. Kuzmichev, T. E. Shanygina, S. N. Tchesnokov, S. I. Krasnosvobodtsev, Solid State Commun. \textbf{152}, 119 (2012).
\bibitem{MgB2 2} S. A. Kuzmichev, T. E. Kuzmicheva, and S. N. Tchesnokov, JETP Letters \textbf{99}, 295 (2014) [Pisma ZheTF {\bf 99}, 339 (2014)].
\bibitem{MB2004JetpL} Ya.G. Ponomarev, S.A. Kuzmichev, N.M. Kadomtseva, M.G. Mikheev, M.V. Sudakova, S.N. Chesnokov, E.G. Maximov, S.I. Krasnosvobodtsev, L.G. Sevast'yanova, K.P. Burdina, and B.M. Bulychev, JETP Lett. \textbf{79}, 484 (2004) [Pisma ZheTF 79, 597 (2004)].
\bibitem{MB2004SSC} Ya.G. Ponomarev, S.A. Kuzmichev, M.G. Mikheev, M.V. Sudakova, S.N. Tchesnokov, N.Z. Timergaleev, A.V. Yarigin, E.G. Maksimov, S.I. Krasnosvobodtsev, A.V. Varlashkin, M.A. Hein, G. M\"{u}ller, H. Piel, L.G. Sevastyanova, O.V. Kravchenko, K.P. Burdina, B.M. Bulychev, Solid State Comm. \textbf{129}, 85 (2004).
\bibitem{EPL} T.E. Kuzmicheva,  S.A. Kuzmichev, M.G. Mikheev, Ya.G. Ponomarev, S.N. Tchesnokov, Yu.F. Eltsev, V.M. Pudalov, K.S. Pervakov, A.V. Sadakov, A.S. Usoltsev, E.P. Khlybov, and L.F. Kulikova, Europhys. Lett. \textbf{102}, 67006 (2013).
\bibitem{SmPRB} T.E. Kuzmicheva, S.A. Kuzmichev, K.S. Pervakov, V.M. Pudalov, N.D. Zhigadlo, Phys. Rev. B \textbf{95}, 094507 (2017).
\bibitem{Nd} T.E. Kuzmicheva, S.A. Kuzmichev, N.D. Zhigadlo, Phys. Rev. B \textbf{100}, 144504 (2019).
\bibitem{LOFA} Ya.G. Ponomarev, S.A. Kuzmichev, M.G. Mikheev, M.V. Sudakova, S.N. Tchesnokov, O.S. Volkova, A.N. Vasiliev, T. H\"{a}nke, C. Hess, G. Behr, R. Klingeler, and B. B\"{u}chner, Phys. Rev. B \textbf{79}, 224517 (2009).

\end{thebibliography}

\end{document}